\documentstyle[12pt]{article}
\makeatletter
\newcommand{\singlespacing}{\let\CS=\@currsize\renewcommand{\baselinestretch}
{1.0}\tiny\CS}
\newcommand{\doublespacing}{\let\CS=\@currsize\renewcommand{\baselinestretch}
{1.5}\tiny\CS}
\begin{document}
\newcommand{\bd}{\begin{document}}
\newcommand{\ed}{\end{document}}
\newcommand{\bc}{\begin{center}}
\newcommand{\ec}{\end{center}}
\newcommand{\bfr}{\begin{flushright}}
\newcommand{\efr}{\end{flushright}}
\newcommand{\lt}{\left}
\newcommand{\rt}{\right}
\newcommand{\vs}{\vspace}
\newcommand{\hs}{\hspace}
\newcommand{\beq}{\begin{equation}}
\newcommand{\eeq}{\end{equation}}
\newcommand{\lb}{\linebreak}
\newcommand{\pb}{\pagebreak}
\newcommand{\mb}{\makebox}
\newcommand{\fb}{\framebox}
\newcommand{\mc}{\multicolumn}
\newcommand{\ben}{\begin{enumerate}}
\newcommand{\een}{\end{enumerate}}
\newcommand{\bit}{\begin{itemize}}
\newcommand{\eit}{\end{itemize}}
\newcommand{\ol}{\overline}
\newcommand{\un}{\underline}
\newcommand{\lefq}{\lefteqn}
\newcommand{\ba}{\begin{array}}
\newcommand{\ea}{\end{array}}
\newcommand{\beqa}{\begin{eqnarray}}
\newcommand{\eeqa}{\end{eqnarray}}
\newcommand{\beqas}{\begin{eqnarray*}}
\newcommand{\eeqas}{\end{eqnarray*}}
\newcommand{\bfg}{\begin{figure}}
\newcommand{\efg}{\end{figure}}
\newcommand{\bds}{\begin{displaymath}}
\newcommand{\eds}{\end{displaymath}}
\newcommand{\btb}{\begin{tabbing}}
\newcommand{\etb}{\end{tabbing}}
\newcommand{\para}{\parallel}
\newcommand{\pad}{\partial}
\newcommand{\nn}{\nonumber}
\newcommand{\la}{\leftarrow}
\newcommand{\ra}{\rightarrow}
\newcommand{\lgla}{\longleftarrow}
\newcommand{\lgra}{\longrightarrow}
\newcommand{\La}{\Leftarrow}
\newcommand{\Ra}{\Rightarrow}
\newcommand{\Lra}{\Leftrightarrow}
\newcommand{\Lgla}{\Longleftarrow}
\newcommand{\Lgra}{\Longrightarrow}
\newcommand{\bm}{\boldmath}
\newcommand{\lan}{\langle}
\newcommand{\ran}{\rangle}
\renewcommand{\a}{\alpha}
\renewcommand{\b}{\beta}
\newcommand{\g}{\gamma}
\newcommand{\G}{\Gamma}
\renewcommand{\d}{\delta}
\newcommand{\eps}{\epsilon}
\newcommand{\th}{\theta}
\newcommand{\Th}{\Theta}
\newcommand{\s}{\sigma}
\newcommand{\lam}{\lambda}
\newcommand{\D}{\Delta}
\newcommand{\vare}{\varepsilon}
\newcommand{\pr}{\prime}
\newcommand{\ro}{\rho}
\newcommand{\nab}{\nabla}
\newcommand{\m}{\mu}
\newcommand{\n}{\nu}
\newcommand{\Sg}{\Sigma}
\newcommand{\p}{\pi}
\newcommand{\R}{I\!\!R}
\newcommand{\om}{\omega}
\newcommand{\Om}{\Omega}
\newcommand{\ze}{\zeta}
\newcommand{\vart}{\vartheta}
\newcommand{\tri}{\triangle}
\newcommand{\f}{\frac}
\newcommand{\iny}{\infty}
\newcommand{\pro}{\propto}

\bc
{\Large \bf  ${ \cal{PT}}$-invariant one-dimensional Coulomb problem }
\ec

\vs{1cm}

\bc
{\large \bf Anjana Sinha $^*$} \\
{\it Department of Applied Mathematics}\\
{\it University of Calcutta}\\
{\it 92 A.P.C.Road, Kolkata - 700 009}\\
{\it INDIA} 

\vs{.5cm}

{\it and} \\

\vs{.5cm}

{\large \bf Rajkumar Roychoudhury $^{\$}$}\\
{\it Physics \& Applied Mathematics Unit}\\
{\it Indian Statistical Institute}\\
{\it Kolkata - 700 108} \\
{\it INDIA}

\ec


\vs{3cm}

\noindent
-----------------------------------------------------------
------------------------------------- \\
e-mail : \\
$^*$ a.sinha@cucc.ernet.in, anjana23@rediffmail.com \\
$^{\$}$ raj@isical.ac.in \\

\pb

\bc
{\bf \large \un{Abstract}}
\ec

The one-dimensional Coulomb-like potential with a real coupling constant 
$ \beta$, and a centrifugal-like core of strength $ G = \alpha ^2 - \f{1}{4} $,
viz., $ V(x) = \f{\alpha ^2 - \f{1}{4}}{(x-ic )^2} ~+~ 
\f{\beta}{\vert x - ic \vert } $ ,
is discussed in the framework of $\cal{PT}$-symmetry. The
$\cal{PT}$-invariant exactly solvable model so formed, is found to 
admit a double set of real and discrete energies, numbered by a 
quasi-parity $q = \pm 1$.

\vspace{5cm}

\noindent
-----------------------------------------------------------
------------------------------------- \\

\noindent
{\bf \un{Key words :}} $\cal{PT}$- symmetry, non-Hermitian Hamiltonian, 
Coulomb potential.

\newpage

\section*{I. Introduction}

Ever since it was conjectured by  Besis, Bender and Boettcher, 
and others [1-3] that Hermiticity of the Hamiltonian is not essential for
the  reality of the spectrum, non-Hermitian Hamiltonians have attracted a lot
of attention. The main reason for this is that a large number of
one-dimensional complex potentials, invariant under the simultaneous actions
of space- and time-reflection operators $ \cal{P}$ and $\cal{T}$ respectively,
have been found to admit real and discrete energies. Though this unusual 
behaviour was at first thought to be because of the so-called 
$\cal{PT}$-invariance, this condition is neither necessary nor 
sufficient to ensure the reality of the spectrum. Some authors have pointed
out that the necessary, but not sufficient, condition for the spectrum to be
real and discrete is the $\eta$-pseudo-Hermiticity, $ \eta {\cal{H}} \eta ^{-1} 
= {\cal{H}} ^{+} $, where $ \eta$ is a Hermitian linear automorphism [4-7]. 
However, no general condition has been found for the breakdown of 
$\cal{PT}$-symmetry either. Several authors have studied many 
one-dimensional non-Hermitian models, constrained by their 
$\cal{PT}$-invariance [2-12], and shown that such 
Hamiltonians exhibit 2 types of behaviour --- 

\vspace{0.2cm}

\noindent
(i) ~~ In the unbroken ${ \cal{PT}}$-symmetry phase, the eigenfunctions of 
the Hamiltonian ${\cal{H}}$ are also eigenfunctions of 
${ \cal{PT}}$, and the energy spectrum is real and discrete. 

\vspace{0.2cm}

\noindent
(ii) ~~ When  ${ \cal{PT}}$-symmetry is spontaneously broken, 
though the potential retains ${ \cal{PT}}$-symmetry, 
the corresponding wavefunctions cease to be the eigenfunctions of 
the operator $\cal{PT}$, and the energy eigenvalues 
are arranged as complex conjugate pairs.

The interesting fact observed about ${\cal{PT}}$-symmetry is that 
the non-Hermitian, ${\cal{PT}}$-invariant models 
admit some of the properties of the usual
Hermitian ones, {\it viz.}, supersymmetry, potential algebra,
quasi-solvability, etc. This has motivated many
authors to study such non-Hermitian Hamiltonians, especially 
because of their applications in many areas of theoretical physics --- 
nuclear physics, quantum field theories, scattering problems, 
localization-delocalization transitions in superconductors, 
defraction of atoms by standing light waves, as well as the
study of solitons on a complex Toda lattice [13].

In this note we study the $\cal{PT}$-symmetric one-dimensional
Coulomb-like potential, with a centrifugal-like core, at a real coupling 
$ \beta $ :
\beq
V(x) = \f{\alpha ^2 - \f{1}{4}}{(x-ic )^2} ~+~ \f{\beta}{\vert x - ic \vert } 
\eeq
The motivation for investigating such a system arises from 
the fact  that the Harmonic oscillator and the 
Coulomb potentials are the most widely studied of all quantum 
mechanical systems. Though the $\cal{PT}$-symmetric version of the
one-dimensional Harmonic oscillator was explored by Znojil [14] 
three years ago, the one-dimensional analogue of the Coulomb potential is yet
to be investigated. It is worth mentioning here that Znojil 
has regularized the Coulomb potential at an imaginary coupling, and developed
it from the Harmonic oscillator by the $\cal{PT}$-symmetric 
Kustaanheimo-Steifel (KS) transformation [15]. However, 
the KS-type mappings also
change the dimensions and angular momenta and the energies of one system
are related to the coupling constants of the other and vice versa. 
Thus the Harmonic oscillator in $D$ dimension is mapped into the 
$d$-dimensional Coulomb problem.
The approach used in the present work is totally different. The Coulomb
potential is treated in one dimension, with a real coupling constant.

The organization of the study is as follows. To make the work self-contained,
a brief review of ${ \cal{PT}}$-symmetry is given in Section II. 
In Section III,  the one-dimensional Coulomb problem is developed in the 
framework of  $\cal{PT}$-symmetric quantum mechanics.
The wave functions are normalized in Section IV.
Section V is kept for conclusions and discussions.

\vs{2cm}

\section*{II. ${ \cal{PT}}$-symmetry}

Before proceeding further, let us briefly recapitulate the idea of 
$\cal{PT}$-symmetry. The  Hamiltonian ${\cal{H}}$ 
for a particle of mass $m$, in a complex 
potential $ V(x) = V_R (x) + i V_I (x) $ is given by
\beq
{ \cal{H}} ~=~ - \f{1}{2m} \f{d^2}{dx^2} ~+~ V(x) 
\eeq
${ \cal{H}}$ is said to be ${ \cal{PT}}$-symmetric when
\beq
{ \cal{PTH}} ~=~ { \cal{HPT}}
\eeq
Here ${ \cal{P}}$ is the {\it Parity operator} acting as spatial reflection,
and ${ \cal{T}}$ stands for {\it Time Reversal}, acting as the complex
conjugation operator. Their action on the position and momentum operators are
given by :
$${ \cal{P}} ~:~ x \ra -x, ~~~p \ra -p $$
$${ \cal{T}} ~:~ x \ra x, ~~~~p \ra -p, ~~~~ i \ra -i $$ 
Hence, in explicit form, the condition for a potential to be 
${\cal{PT}}$-symmetric is 
\beq
\lt [ V(-x) \rt ] ^* ~=~ V(x)
\eeq
so that $V_R(x)$ must be an even function of $x$, whereas $V_I(x)$ has to be
odd. The commutation relation 
\beq
\lt [ x, p \rt ] ~=~ i \hbar 
\eeq
remains invariant under ${ \cal{PT}}$ for both real as well as complex $x$ and
$p$. 

For the probabilistic interpretation of the wave function of such a
non-Hermitian quantum mechanical system, the norm and the inner product have
to be redefined [16,17]. The counterpart of the scalar product of two
eigenfunctions $\psi _1 (x)$ and $\psi _2 (x)$ is defined as
\beq
< \psi _2 \vert  \psi _1 > ~=~ \int ~dx ~\psi _2 ^* (-x) ~ \psi _1 (x) 
\eeq
and the normalization condition is replaced by
\beq
\int _{- \infty} ^{\infty} ~dx ~\psi ^* (-x) ~ \psi (x) ~=~ \sigma
\ \ \ \ , 
\ \ \ \ \ \ \ \ \ \ \ \  \sigma ~=~ \pm 1
\eeq
It is evident that the norm is not necessarily positive-definite. So it is
referred to as the pseudo-norm. $\sigma = 1$ corresponds to the
$\cal{PT}$-symmetric phase while $\sigma = - 1$ corresponds to the
$\cal{PT}$-antisymmetric phase.

\vs{2cm}

\section*{III. One-dimensional Coulomb-like potential}

We start with the radial Coulomb problem in three dimensions.
The Schr\"{o}dinger equation for this system (in units $\hbar = 2m = 1$) is
\beq	
-\f{d^2 \psi}{dr^2} ~+~ \lt \{  \f{ l(l+1)}{r^2} ~-~ 
\f{1}{r} \rt \} \psi  ~=~   E \psi
\eeq
Changing its variable  $r$ to $ x = r + ic $, $ \  c>0$, and replacing the
angular momentum $l$ by $ \alpha - \f{1}{2} $ [14], we obatin the 
$\cal{PT}$-invariant one-dimensional
Coulomb-like potential, with a centrifugal-like core of strength 
$ G ~=~ \alpha ^2 ~-~ \f{1}{4} $ 
\beq
V(x) = \f{\alpha ^2 - \f{1}{4}}{(x-ic )^2} ~+~ \f{\beta}{\vert x - ic \vert } 
\eeq
Here $ \beta $ is a real coupling constant and $ \alpha > 0 $. 
The analytic 
continuation of the model to one dimension makes the boundary 
condition at the origin redundant. The shift $c$ avoids the 
singularity from the integration path, rendering the Hamiltonian exactly
solvable on the entire line for any value of $ \alpha > 0 $.  
The Schr\"{o}dinger equation (8) then takes the form
\beq	
\f{d^2 \psi}{dx^2} ~+~ \lt \{ E ~-~ \f{ \alpha ^2 - \f{1}{4}}{(x-ic)^2} ~-~
\f{\beta}{|x-ic|} \rt \} \psi  ~=~ 0  
\eeq
After some straightforward algebra, the solution to eq. (10) is obtained as
\beq
\psi (x) ~=~ N_{qn} ^ {\alpha} e^{- \f{ \gamma}{2} \vert x - ic \vert }~ 
\lt \{ \gamma \vert x - ic \vert \rt \} 
^{- q \alpha + \f{1}{2} } ~ L_n ^{-2q \alpha} \lt \{ \gamma \vert
x - ic \vert \rt \} 
\eeq
where  $ N_{qn} ^{\alpha}$ is the normalization coefficient 
\beq
N_{qn} ^{\alpha} ~=~ \vert N_{qn} ^{\alpha} \vert e^{i \nu}
\eeq 
\beq
\gamma = 2 \sqrt{ | E _{qn} ^ { \alpha }| }
\eeq 
and $ L_n ^{ -2q \alpha }$ are the associated Laguerre polynomial [18]
\beq
L_n ^m (x) = \sum _{j=0} ^n (-1) ^j \lt (
\begin{array}{c} n+m \\ n-j \\ \end{array}  \rt ) \f{x ^j}{j !}
\eeq
\beq
L_0 ^m (x) = 1
\eeq
\beq
L_1 ^m (x) = m+1 -x
\eeq
From eq. (11),  \ \  $ \psi ^{*} (-x) ~=~ \psi (x) $,  \ \ so
that this is a case of unbroken $\cal{PT}$-symmetry. As expected, 
the spectrum turns out to be real and discrete. The eigenenergies 
are obtained as a double series, parametrized 
by a quantum number $ q = \pm 1 $, given by
\beq
E_{qn} ^{\alpha} ~=~ - \f{ \beta ^2}{ \lt (2n - 2q \alpha + 1 \rt )^2 }
\eeq
$q$ may be called the {\it quasi-parity} of the system. In the limit 
$ c \ra 0 $, the Hamiltonian gets back
its hermiticity, reverting back to the ordinary radial Coulomb problem, 
and the quasi-parity degenerates to the ordinary parity.

\vs{2cm}

\section*{IV. Normalization of the wave function}

In $\cal{PT}$-symmetric quantum mechanics, $ \psi (x)$ and $ \psi ^* (-x)$
both satisfy the time-independent Schr\"{o}dinger equation 
\beq
- \f{d^2 \psi }{dx^2} + V(x) \psi ~=~ E \psi 
\eeq
If $E$ is real and non-degenerate, then 
\beq
\psi ^* (-x) ~=~ e ^{i \phi} \psi (x) \ \ \ \ \ \ \ \ , 
\ \ \ \ \ \   0 \leq \phi < 2 \pi
\eeq
For the phase factor $ \phi = 0 $, $ \psi (x) $ is $ \cal{PT}$-symmetric,
and for $ \phi = \pi $, $ \psi (x) $ is $ \cal{PT}$-antisymmetric . \\
For the particular potential we are dealing with, $ \phi = 0 $, so that the
functions $ \psi (x)$ are $ \cal{PT}$-symmetric. Therefore the normalization
condition reads
\beq
\int _{- \infty} ^{\infty} ~dx ~\psi ^* (-x) ~ \psi (x) ~=~ 1
\eeq
Now
\beq
\int _{- \infty} ^{\infty} ~dx ~\psi ^* (-x) ~ \psi (x) ~=~ \vert N_{qn} 
^ {\alpha} \vert ^2 ~ I_{qn} ^{\alpha}
\eeq
where
\beq
I_{qn} ^{\alpha} ~=~ \int _{ - \infty} ^{ \infty} dx ~ ~ 
e^{-  \gamma \vert x - ic \vert }~ \lt \{ \gamma \vert x - ic \vert \rt \} 
^{- 2q \alpha + 1} ~ \lt \{ L_n ^{-2q \alpha} \lt ( \gamma 
\vert x - ic \vert \rt ) \rt \} ^2
\eeq
$ I_{qn} ^{\alpha}$ is evaluated by performing the contour integration 
$ \int dz ~  f(z) $ in terms of the complex variable $z = |x-ic|$, 
with $f(z) $ being given by 
\beq
f(z)  ~=~ \int _{ - \infty} ^{ \infty} dz ~ 
e^{-  \gamma z }~ \lt \{ \gamma z \rt \} ^{- 2q \alpha + 1} ~ 
\lt \{ L_n ^{-2q \alpha} \lt [ \gamma z \rt ] \rt \} ^2
\eeq

\setlength{\unitlength}{1.5mm}
\begin{picture}(90,28)
\put(0,0){\vector(1,0){12}}
\put(12,0){\line(1,0){18}}
\put(-2,-2){\makebox(0,0){\rm P}}
\put(30,0){\vector(1,0){10}}
\put(40,0){\line(1,0){20}}
\put(62,-2){\makebox(0,0){\rm Q}}
\put(28,-1){\makebox(0,0){\rm c}}
\put(40,-2){\makebox(0,0){\rm $ \Gamma $}}
\put(60,0){\vector(0,1){6}}
\put(60,6){\line(0,1){6}}
\put(63,11){\makebox(0,0){\rm R}}
\put(30,-10){\vector(0,1){30}}
\put(30,22){\makebox(0,0){\rm Im z}}
\put(-12,13){\vector(1,0){84}}
\put(78,13){\makebox(0,0){\rm Re z}}
\put(0,12){\vector(0,-1){6}}
\put(0,6){\line(0,-1){6}}
\put(-3,11){\makebox(0,0){\rm U}}
\put(62,15){\makebox(0,0){\rm $ + \Lambda $ }}
\put(0,15){\makebox(0,0){\rm $ - \Lambda $}}
\put(60,12){\vector(-1,0){10}}
\put(50,12){\line(-1,0){19}}
\put(33,15){\makebox(0,0){\rm $ + \eps$ }}
\put(29,12){\vector(-1,0){15}}
\put(14,12){\line(-1,0){14}}
\put(27,15){\makebox(0,0){\rm $ - \eps$}}
\put(30,12){\oval(2,2)[b]}
\put(34,10){\makebox(0,0){\rm S}}
\put(26,10){\makebox(0,0){\rm T}}

\put(30,-15){\small {\makebox(0,0){\rm Fig. 1. \ \  Contour $\Gamma$ 
in the complex plane}}}

\end{picture}

\vs{4cm}

\noindent
Thus $ f(z) $ is analytic everywhere except at the branch point $ z = 0 $.
Enclosing the contour $ \Gamma \equiv PQRSTU $ in the complex plane, avoiding
the singularity at $z=0$ by a semicircle of radius $ \eps $, 
one obtains $ \int _{\Gamma} dz ~ f(z) = 0 $ from Cauchy's theorem. 
Detailed calculations show that in the limit $ \Lambda \ra \infty $
and $ \eps \ra 0 $, \  \ 
$ \int _ {- \infty} ^ { \infty} dz ~ f(z) ~\ra~ 2 ~ \int _ 0 ^ 
{\infty} dx ~ f(x)$. Also, $ I_{qn}   ^{\alpha} $ converges for 
$ q \alpha < 1 $. Thus for $ q = \pm 1$, $ \alpha $ should be restricted 
in the range $ 0 < \alpha < 1$.

\noindent
After some algebra, $ I_{qn} ^{\alpha}$ is calculated to be [18]
$$ I_{qn} ^{\alpha} ~=~  \f{2(-2q \alpha +1)}{\gamma} ~ 
\f{ \Gamma (-2q \alpha + n + 1) ~ }{ (n! )^2 } $$
\beq
. \lt \{ \f{d^n}{dh ^n} \lt [ \f{ (1-h) ~ F \lt ( -q \alpha +1, ~ -q \alpha +
\f{3}{2}, ~ -2q \alpha +1, ~ \f{4h}{(1+h)^2} \rt ) }{  ( 1 + h) 
^{-2q \alpha+2} }  \rt ] \rt \} _{h=0}
\eeq
where $F$ are the generalised Hypergeometric functions [18]. Thus 
the normalization coefficient is evaluated from 
\beq
\vert N_{qn} ^{\alpha} \vert ~=~ \lt [ \f{1}{ I_{qn} ^{\alpha}  } 
\rt ] ^{~ \f{1}{2}}
\eeq
It is easy to observe that for the ground state ,
\beq
\vert N_{q0} ^{\alpha} \vert ~=~ \lt \{ \f{ \sqrt{ E_{q0} ^{\alpha}} }{ 
\Gamma ( -2q \alpha+2 )  } \rt \} ^{\f{1}{2}}
\eeq
and for the first excited state,
\beq
\vert N_{q1} ^{\alpha} \vert ~=~ \lt \{ \f{ \sqrt{ E_{q1} ^{\alpha}} }{ 
\lt ( -2q \alpha+3 \rt ) ~ \Gamma (-2q \alpha + 2) } \rt \} ^{\f{1}{2}}
\eeq

\vs{2cm}

\section*{V. Conclusions and discussions}

The $\cal{PT}$-invariant one-dimensional Coulomb-like potential 
$ \f{\beta}{|x-ic|}$, together with a centrifugal core 
$ \f{ \alpha ^2 - 1/4}{ (x - ic )^2} $, has been solved exactly at a real
coupling constant $ \beta $. In this respect the present analysis is different
from the imaginary coupling regularization of ref. [15]. The shift 
$c$ avoids the singularity from the integration path. In the limit 
$c \ra 0 $, the Hamiltonian reduces to that of the conventional
Coulomb potential, and the quasi-parity degenerates to the ordinary parity.

It is observed that, analogous to 
the $\cal{PT}$-invariant harmonic oscillator [14], the 
eigenfunctions as well as the eigenenergies 
are numbered by a quasi-parity $ q = \pm 1$, 
and the spectrum is found to be real and bounded. Thus the
$\cal{PT}$-symmetric version gives rise to a second set of bound states. 
Moreover, these energies are found to depend on the value of $ \alpha $. 
$ \alpha = 0 $ is not allowed as the quasi-parity no longer exists in this
case. Normalization of the wave functions applying the modified 
normalization for
$\cal{PT}$-symmetry, restricts the value $ q \alpha < 1$, 
which for $ q = \pm 1$, implies $ 0 < \alpha < 1 $. While 
$ q = -1 $ reproduces the ordinary three-dimensional Coulomb energies, viz.,
\beq
E_{n-} ^{ \alpha} = - \f{ \beta ^2 }{ \lt ( 2n + 2 \alpha + 1 \rt ) ^2 } \ \ \
, \ \ \ \ \ n = 0, 1, 2, \cdots
\eeq
$q = +1$ sector further enriches the spectrum. 
For this particular case, 
$ \alpha = \f{1}{2} $ is not allowed for $n=0$, as then 
the energy disappears from the system. Therefore, 
\beq
E_{n+} ^{ \alpha} = - \f{ \beta ^2 }{ \lt ( 2n - 2 \alpha + 1 \rt ) ^2 } \ \ \
, \ \ \ \ \ n = 1, 2, 3, \cdots
\eeq
In a way this may be compared with supersymmetric quantum mechanics, where the
partner potentials are isospectral, with the possible exception of the ground
state. 

To conclude, the analysis in the present work is different from the
imaginary coupling regularization in ref. [15], where one encounters
unavoidable level crossings at all positive integral and half-integral values
of $ \alpha$. Moreover, for $ q = + 1 $, there are a plethora of 
{\it flown-away} energies at $ \alpha = n + \f{1}{2} $. However, 
it is shown in this study that $ \alpha $
is restricted to lie in the range $ 0 < \alpha < 1 $, and hence such phenomena
are not observed here. The single flown-away energy at $ \alpha = \f{1}{2} $
for $q = +1 $ can be avoided if one restricts $n$ to $ n = 1, 2, 3, \cdots $.

\vs{2cm}

\section*{Acknowledgment}

The authors are grateful to Prof. B. Bagchi for some enlightening discussions
on the topic. One of the authors (A.S.) acknowledges financial assistance from
the Council of Scientific and Industrial Research, India. 

\pb

\section*{References}

\begin{enumerate}
\item[1.] D. Besis, unpublished (1992).
\item[2.] C. M. Bender \& S. Boettcher, Phys. Rev. Lett. {\bf 80} 5243 (1998),
J. Phys. {\bf A 31} L273 (1998).
\item[3.] C. M. Bender, S. Boettcher \& P. N. Meisinger, J. Math. Phys. 
{\bf 40} 2201 (1999).
\item[4.] A. Mostafazadeh, J. Math. Phys. {\bf 43} 205 (2002), {\bf 43} 
2814 (2002).
\item[5.] Z. Ahmed, Phys. Lett. A : {\bf 282} 343 (2001),
{\bf 290} 19 (2001), {\bf 294} 287 (2002).
\item[6.] T. V. Fityo, arXiv : quant-ph/0204029. 
\item[7.] Chun-Sheng Jia, Pi-Yuan Lin \& Liang-Tian Sun,  Phys. Lett. 
{\bf A } {\it In press} (2002).
\item[8.] G. L\'{e}vai \& M. Znojil, arXiv : quant-ph / 0206013, J. Phys. A 
{\bf 33} 7165 (2000).
\item[9.] F. Cannata, G. Junker \& J. Trost, Phys. Lett. {\bf A 246} 219
(1998). 
\item[10.] B. Bagchi and R. Roychoudhury, J. Phys. A : Math. Gen. {\bf 33}
L1 (2000).
\item[11.] G. L\'{e}vai, F. Cannata and A. Ventura, arXiv : quant-ph / 0206032.
\item[12.] M. Znojil, J. Phys. A {\bf 33} 4561 (2000),
J. Phys. A {\bf 35} 2341 (2002),
Phys. Lett. A {\bf 285} 7 (2001).
\item[13.] A. Sinha \& R. Roychoudhury, Phys. Lett. A {\it In press} (2002), 
{\it and references therein}.
\item[14.] M. Znojil, Phys. Lett. A {\bf 259} 220 (1999).
\item[15.] M. Znojil and G. L\'{e}vai, Phys. Lett. A {\bf 271} 327 (2000). 
\item[16.] B. Bagchi, C. Quesne, \& M. Znojil, Mod. Phys. Lett. {\bf A 16} 
2047 (2001).
\item[17.] G. S. Japaridze, J. Phys. {\bf A 35} 1709 (2002).
\item[18.] I. S. Gradshteyn \& I. M. Ryzhik, Tables of Integrals, Series and
Products, AP, New York, (1980).
\end{enumerate}

\end{document}